\documentclass[unsortedaddress,onecolumn,showpacs,showkeys]{revtex4}

\usepackage{amsmath}
\usepackage{amsfonts}
\renewcommand{\Re}{\operatorname{Re}}
\newcommand{\pcd}{probability current density}
\newcommand{\schro}{Schr\"odinger equation}

\begin{document}
\title{Resonant tunneling with zero reflection at the classical velocity}
\author{C. Pacher}
\affiliation{Austrian Research Centers GmbH - ARC, Smart Systems Division, Donau-City Stra\ss e 1, A-1220 Wien, Austria}
\date{\today}

\begin{abstract}
An important aspect of resonant tunneling with a probability of unity (thus zero reflection) through a finite region with length $l$ is studied. The relation between the velocity expectation value $\langle \hat v_{\textrm{res}}\rangle$ restricted to a region of length $l$ and the tunneling time $\tau_{\textrm{res}}$ through the same region is calculated. The obtained result is the analogue of the mean velocity in classical mechanics: The velocity expectation value equals exactly the length divided by the tunneling time. This result holds for any potential but is especially relevant for finite periodic potentials and inversion symmetric potentials where resonances show a tunneling probability of unity.
\end{abstract}
\pacs{03.65.Xp,03.65.-w}
\keywords{tunneling time, velocity, resonant tunneling} 
\maketitle

\section{Introduction}
The calculation of an interaction time or tunneling time associated with the tunneling process through a classically forbidden region is a quite controversial topic. Most of the studies have concentrated on electron tunneling with non-zero reflection through a single rectangular barrier. Often, at least in intermediate steps, complex times resulted from these analyses. See Refs.~\onlinecite{Hauge89,Landauer94,Carvallo02} for recent reviews of the different definitions and argumentations.

The phenomenon of resonant tunneling - through a classically forbidden region - with a transmission probability $|T|^2$ (where $T$ is the complex transmission amplitude) that reaches unity seems to be easier to understand than tunneling with a transmission below unity. For energies where perfect transmission, i.e. $|T|^2=1$, occurs the following relations hold additionally: the complex reflection coefficient $R=0$, $\partial |T|/\partial E = 0$, and the incoming and transmitted probability current densities are identical. 
Under these conditions all tunneling related times $\tau$ of which the author is aware of [Bohm-Wigner time (phase delay time)\cite{Bohm51,Wigner55}, B\"uttiker-Landauer time\cite{Buettiker82}, Larmor time\cite{Buettiker83}, Pollak-Miller time\cite{Pollak84}, dwell time\cite{Smith60,Buettiker83}, the complex traversal time by Sokolovski and Baskin\cite{Sokolovski87} and the tunneling time introduced by Li and Wang\cite{Li2001}] are identical and real-valued. Except for the last mentioned time (which can be easily seen to be in the resonant case identical to the dwell time), the identity between the other times has been discussed in Ref.~\onlinecite{Pacher05}. In Ref.~\onlinecite{Pacher05} we also calculated the velocity and the tunneling time in one-dimensional \emph{finite periodic systems} (FPS) at transmission resonances. We observed that the expectation value of the velocity (restricted to the finite system) fulfills exactly the classical relation $\langle \hat v_{\textrm{res}}^{\textrm{FPS}}\rangle = l/\tau_{\textrm{res}}^{\textrm{FPS}}$.

Based on this work\cite{Pacher05} here we will give a short self contained proof that the classical relation $\langle \hat v_{\textrm{res}}\rangle = l/\tau_{\textrm{res}}$ must indeed hold for an \emph{arbitrary} one-dimensional system at perfect transmission resonances \cite{Differentiate}.

\section{Theory}
The Hamiltonian of our system is $\hat H=\hat p^2/2m +V(x)$, where the potential is zero outside the region $0\le x \le l$, i.e.
\begin{equation}
V(x)=\left\{
\begin{array}
[c]{ll}%
0 &  x < 0,\\
\tilde V(x) & 0\le x\le l,\\
0 & x > l.
\end{array}
\right.
\end{equation}

We restrict ourselves to the study of stationary solutions in the form of plane waves (scattering solutions). In this case the solution of the time-dependent \schro, $\hat H\Psi(x,t)=i\hbar\frac{\partial}{\partial t} \Psi(x,t)$, is given by
\begin{equation}
\Psi(x,t)=\exp(-iEt/\hbar) \times \left\{
\begin{array}
[c]{ll}%
\exp(ikx)+R\exp(-ikx)\quad &  x\le0,\\
\Phi(x) & 0\le x\le l,\\
T\exp[ik(x-l)] & x\ge l.
\end{array}
\right.  \label{eq:PsiGen}
\end{equation}
In the case of perfect transmission we have $R=0$ and $T=\exp(i\alpha)$ with $\alpha\in \mathbb{R}$ and get
\begin{equation}
\Psi(x,t)=\exp(-iEt/\hbar) \times \left\{
\begin{array}
[c]{ll}%
\exp(ikx)\quad &  x\le0,\\
\Phi(x) & 0\le x\le l,\\
\exp(i\alpha)\exp[ik(x-l)] & x\ge l.
\end{array}
\right.  \label{eq:PsiAtRes}
\end{equation} 
The probability current density is given by
\begin{equation}
j(x,t)=\Re\left\{%
\Psi^*(x,t)\left[-\frac{i\hbar}{m}\frac{d}{dx}\Psi(x,t)\right]%
\right\}. \label{eq:probCurr}
\end{equation} 

In one dimension the continuity equation reads
\begin{equation}
\frac{\partial|\Psi(x,t)|^2}{\partial t} + \frac{\partial j(x,t)}{\partial x}=0. 
\label{eq:contEq}
\end{equation} 
Of course, for a stationary state $j(x,t)$ and $|\Psi(x,t)|^2$ do not depend on time. 
Equation (\ref{eq:contEq}) shows that in one dimension the probability current of a stationary state does not depend on space as well, i.e. $j(x,t)=j_{0}$. Up to now, everything is well established.


Integrating the identity $j(x,t)=j_0$ between $x=0$ and $x=l$ and inserting Eqs.~(\ref{eq:probCurr}) and (\ref{eq:PsiAtRes}) we obtain
\begin{equation}
lj_0=\int_0^l dx j(x,t)=\Re\int_0^l dx
\Phi^*(x)\left[-\frac{i\hbar}{m}\frac{d}{dx}\Phi(x)%
\right]. \label{eq:intj0}
\end{equation} 

For perfect transmission we have 
\begin{equation}
|T|=1\Rightarrow j_{0}=j_{in}, \label{eq:j0equalsjin}
\end{equation}
where $j_{in}$ denotes the \pcd\ of the incoming wave \mbox{$\exp(ikx-iEt/\hbar)$} and the integral in Eq.~(\ref{eq:intj0}) is real-valued, i.e.
\begin{equation}
|T|=1\Rightarrow \int_0^l dx \Phi^*(x)\left[-\frac{i\hbar}{m}\frac{d}{dx}\Phi(x)\right] \in \mathbb{R}.
\label{eq:intIsReal}
\end{equation}
The real-valuedness of the integral can easily be shown by partial integration (explicitely done in the appendix). 
As a consequence of Eqs.~(\ref{eq:intj0})-(\ref{eq:intIsReal}) we obtain at resonance
\begin{equation}
|T|=1 \Rightarrow l=\frac{1}{j_{in}}\int_0^l dx
\Phi^*(x)\left[-\frac{i\hbar}{m}\frac{d}{dx}\Phi(x)%
\right]. \label{eq:intj0_re}
\end{equation} 

Since at resonance all different tunneling times are identical, we can choose the dwell time\cite{Buettiker83} $\tau_D$ for our further calculations, i.e.
\begin{equation}
|T|=1 \Rightarrow \tau_{\textrm{res}}=\tau_D:=\frac{1}{j_{in}}\int_0^l dx|\Phi(x)|^2\in\mathbb{R}. \label{eq:tauDwell}
\end{equation}

Now Eqs.~(\ref{eq:tauDwell}) and (\ref{eq:intj0_re}) combined give finally our \emph{exact} result:
\begin{equation}
|T|=1 \Rightarrow \frac{l}{\tau_{\textrm{res}}}=\langle \hat v_{\textrm{res}} \rangle_{0,l}:=\frac{\langle \Phi | \hat v \hat P(0,l)| \Phi \rangle}{\langle \Phi |\hat P(0,l)| \Phi \rangle}\in\mathbb{R},
\end{equation} 
where $\hat v=\hat p/m=-\frac{i\hbar}{m} \frac{d}{dx}$ and 
\begin{equation}
\hat P(0,l)=\int_0^l dx |x\rangle \langle x|
\end{equation} 
denotes the projection operator onto the region $0 \le x \le l$. 

Thus, we arrived at a perfect analogue of the definition of the classical mean velocity, i.e. $\bar{v}=\Delta l/\Delta t$. 

Additionally, this is a novel method to calculate the tunneling time (valid only at a resonance with $|T|=1$) when only the wave function is known:
\begin{equation}
\tau_{\textrm{res}}=\frac{l}{\langle \hat v_{\textrm{res}} \rangle_{0,l}}.
\end{equation} 


\section{Discussion}
Given that the quantum mechanical system that we considered behaves in many respects like a classical one, it is not too surprising that we end up with a classical relation. On the other hand, the result is by no means a trivial one since our derivation is completely built upon quantum mechanics (e.g. in the final equation the expectation value involves the complex wavefunction). As can be seen from the derivation, the important point is indeed that no reflection occurs. This fact is the reason (i) for having a clear definition for the tunneling time, (ii) that the velocity expectation value is real-valued, and (iii) that the probability current through the system is equal to the incoming probability current.

Among the potentials that feature a perfect transmission at some energies (but not restricted to) are all finite (or locally) periodic potentials [$V(x+d)=V(x)$] which have been studied with respect to the current topic in Ref.~\onlinecite{Pacher05} and potentials with inversion symmetry [$V(x)=V(-x)$]\cite{Price93}. 

Possibly, the widest range of application can be found within resonant tunneling of electrons through semiconductor heterostructures. Of course, resonant tunneling of other massive particles (e.g. resonant neutron tunneling, see Refs.~\onlinecite{Steinhauser80,Maaza96,Hino99}, and resonant atom tunneling, see Refs.~\onlinecite{Tribe96,Santos98}) is covered as well.

\section{Conclusion}
An important aspect of resonant tunneling with a probability of unity through a finite region with length $l$ has been studied. Starting from the one-dimensional Schr\"odinger equation we calculated the relation between the velocity expectation value $\langle \hat v_{\textrm{res}}\rangle$ restricted to the region $0\le x \le l$ and the tunneling time $\tau_{\textrm{res}}$ through the same region. The analogue of the mean velocity in classical mechanics is obtained: The velocity expectation value equals exactly the length divided by the tunneling time, i.e. $\langle \hat v_{\textrm{res}} \rangle_{0,l}=\frac{l}{\tau_{\textrm{res}}}$.

The result holds for massive particles (described by the Schr\"odinger equation) in arbitrary potentials under the single condition of a tunneling probability of unity.

\begin{acknowledgments}
It is a pleasure to thank M.~Suda and D.~W.~Sprung for reading the manuscript and for important comments.
\end{acknowledgments}

\begin{appendix}
\section{Proof of Equation~(\ref{eq:intIsReal})}
We have to show that for $|T|=1$ 
\begin{equation}
I:=-\frac{i\hbar}{m}\int_0^l dx \Phi^*(x)\left[\frac{d}{dx}\Phi(x)\right] \in \mathbb{R}
\end{equation} 
holds. Partial integration gives
\begin{equation}
I=-\frac{i\hbar}{m} \left(|\Phi(l)|^2-|\Phi(0)|^2 \right)+ \frac{i\hbar}{m}\int_0^l dx \left[\frac{d}{dx}\Phi^*(x)\right]\Phi(x).
\end{equation} 
From Eq.~(\ref{eq:PsiAtRes}) we have $|T|=1\Rightarrow |\Phi(l)|^2=1=|\Phi(0)|^2$. Therefore 
\begin{equation}
|T|=1\Rightarrow I=I^* \Rightarrow I\in \mathbb{R}.
\end{equation} 
\end{appendix}
\bibliography{../refs,tvcomment}
\end{document}